\documentstyle[12pt]{article}

\begin{document}
\setcounter{page}{1} \pagestyle{plain} \vspace{1cm}
\begin{center}
\Large{\bf Phantom-Like Behavior in Modified Teleparallel Gravity}\\
\small \vspace{1cm} {\bf Sakineh Karimzadeh}\footnote{karimzade.f@gmail.com}$^{*}$ \quad\
and \quad {\bf Raheleh Shojaee}\footnote{r.shojaee@azaruniv.ac.ir}$^{\dag}$\\
\vspace{0.5cm} {\it $^{*}$Department of Physics, Faculty of Basic Sciences,\\
University of Mazandaran, P. O. Box 47416-95447, Babolsar, Iran}\\
\vspace{0.5cm} {\it $^{\dag}$Department of Physics, Azarbaijan Shahid Madani University,\\
P. O. Box 53714-161, Tabriz, Iran}\\
\end{center}
\vspace{1.5cm}
\begin{abstract}
The recent data from Planck2018 shows that the equation of state parameter of effective cosmic fluid today is $w_{0}=-1.03\pm 0.03$.
This indicates that it is possible for the universe to be in a phantom dominated era today.
While a phantom field is essentially capable to explain this observation, it suffers
from some serious problems such as instabilities and violation of the null energy condition. So,
it would be interesting to realize this effective phantom behavior
without adopting a phantom field. In this paper we study possible
realization of an effective phantom behavior in modified teleparallel gravity. We show that modified teleparallel gravity is
able essentially to realize an effective phantom-like behavior in the absence of a
phantom field. For this purpose, we choose some observationally
viable $f(T)$ functions and prove that there are some subspaces of the
models parameter space capable of realizing a phantom-like behavior without adopting a phantom field. 
\end{abstract}
\vspace{1.5cm}
\newpage

\section{Introduction}
Different cosmological observations of the last two decades such as
measurements of luminosity distances of Supernova Type Ia (SNIa)
[1,2], the cosmic microwave background (CMB) temperature anisotropy
through the Wilkinson Microwave Anisotropy probe (WMAP) satellite
[3-5], large scale structure [6], the integrated Sachs-Wolfe effect
[7] and more recently the Planck satellite data [8,9] show that the
universe is currently in a positively accelerating phase of
expansion. Since general relativity with ordinary matter and energy
components cannot explain such a novel observation, different
solutions have been proposed to justify this late time cosmic speed
up. There are generally two approaches to explain this observation
by modification of the Einstein's field equations. The first
approach attempts to modify the energy-momentum (matter) part of the
field equations by adding a yet unknown component dubbed dark
energy, and the second approach that tries to modify the geometric
part of the field equations. Scalar fields dark energy models
[10-12] belong to the first category, while modified gravity
theories such as $f(R)$ gravity [13,14], Gauss-Bonnet and $f(G)$
gravity [15-16], massive gravity [17], Lovelock gravity [18],
extra-dimensional gravity [19,20], Teleparallel Equivalent of
General Relativity (TEGR)[21-25] and $f(T)$ gravity [26-28] belong
to the second category. A simple candidate for dark energy proposal
is the cosmological constant [29,30] and the corresponding
cosmological model (the $\Lambda$CDM) has very good agreement with
recent observational data. But this scenario has not a dynamical
behavior and its equation of state parameter stays always at $-1$.
The origin of cosmological constant is not yet well-understood, it
needs a huge amount of fine-tuning and it is impossible to realize
phantom divide line crossing in the pure $\Lambda$CDM model. Because
of these shortcomings with cosmological constant, in recent years
attentions have been drawn toward modified gravity theories and
among the various kinds of modified gravity models, teleparallel
gravity (TEGR) and $f(T)$ gravity have recently obtained a lot of
regard. Teleparallel gravity is completely equivalent with general
relativity at the level of equations and in this theory the
Lagrangian is written in terms of the torsion scalar $T$. Much
similar to the $f(R)$ gravity that one replaces $R$ by $f(R)$ in the
Einstein-Hilbert action of general relativity, $f(T)$ gravity is an
extension of TEGR that replaces $T$ by a generic function of torsion
as $f(T)$ in teleparallel gravity. In this model the Ricci scalar
$R$ is zero since there is no curvature; instead the torsion field
is considered in the framework of Einstein's other (teleparallel)
gravity. In recent years various aspects of teleparallel and
modified teleparallel gravity are studied in details (see Refs.
[31-37] for some  various works in this field). The observational
viability of teleparallel and modified teleparallel gravity has been
studied in Refs. [38,39].

With an \emph{effective phantom-like behavior} one means that the
effective energy density of the cosmic fluid is positive and
increases with time, and in the same time the effective equation of
state parameter stays less than $-1$ [40] . Typically for
realization of such a behavior phantom fields are considered, while
the existence of phantom field causes instabilities and violates the
null energy condition [41-43]. Possible realization of phantom-like
behavior in some cosmological models such as the DGP braneworld
model [44-46] and $f(R)$ gravity are studied [47]. Also possible
existence of a phantom-like phase in teleparallel and also modified
teleparalel gravity theories has been studied in some references
such as Refs. [48-52].
 In this paper, we study possible realization of the
phantom-like behavior in modified teleparallel, $f(T)$, gravity. In
order to achieve this goal, we choose three viable models of $f(T)$
gravity and with appropriate selection of the models parameters, we
show that phantom-like behavior can be realized without any phantom
fields in these models of modified teleparallel gravity. This result
is important because unlike phantom fields, $f(T)$ gravity respects
the null energy condition and more importantly, matter is always
stable in this framework [53].
 The paper is organized as follows. In section 2, we briefly
review field equations in $f(T)$ gravity. In section 3, we choose
three models of $f(T)$ in order to see whether the adopted models have an
effective phantom behavior or not. For this purpose, we plot the effective energy density and
the equation of state parameter versus the redshift $z$, and compare the results with the latest observational data from Planck2018.
Finally, in section 4, we present the summary and conclusion.

\section{Field Equations for Modified Teleparallel Gravity}

\subsection{Field Equations of $f(T)$ Gravity}

In teleparallel (Einstein's other) gravity one needs to define four
orthogonal vector fields named \emph{tetrads} which form a basis for
spacetime manifold. In this framework, the manifold and the
Minkowski metrics are related as follows [53,54]
\begin{equation}
g_{\mu\nu}=\eta_{ij}e_{\mu}^{i}e_{\nu}^{j}
\end{equation}
where the Greek indices run from $0$ to $3$ in coordinate basis of
the manifold while the Latin indices run the same but in the tangent space of the
manifold and $\eta_{ij}=diag(+1,\,-1,\,-1,\,-1)$ is the Minkowski metric.  The connection in
teleparallel gravity, that is, the Weitzenbock connection, is defined as follows
\begin{equation}
\Gamma^{\rho}_{\,\,\,\,\mu\nu}=e^{\rho}_{i}\partial_{\nu}e^{i}_{\mu}\,,
\end{equation}
which gives the spacetime a nonzero torsion but zero curvature in
contrast to general relativity. By this definition the torsion
tensor and its permutations are
\begin{equation}
T^{\rho}_{\,\,\,\,\mu\nu}\equiv
e_{i}^{\rho}(\partial_{\mu}e_{\nu}^{i}-\partial_{\nu}e_{\mu}^{i})
\end{equation}
\begin{equation}
K^{\mu\nu}_{\quad\rho}=-\frac{1}{2}(T^{\mu\nu}_{\quad\rho}
-T^{\nu\mu}_{\quad\rho}-T_{\rho}^{\,\,\,\,\mu\nu})
\end{equation}
\begin{equation}
S^{\,\,\,\,\mu\nu}_{\rho}=\frac{1}{2}(K^{\mu\nu}_{\quad\rho}
+\delta^{\mu}_{\rho}T^{\alpha\nu}_{\quad\alpha}-\delta^{\nu}_{\rho}
T^{\alpha\mu}_{\quad\alpha})\,,
\end{equation}
where $S^{\,\,\,\,\mu\nu}_{\rho}$ is called the \emph{Superpotential}. In
correspondence with Ricci scalar we define a torsion scalar as
\begin{equation}
T=S^{\,\,\,\,\mu\nu}_{\rho}T^{\rho}_{\,\,\,\,\mu\nu}\,.
\end{equation}
So, the gravitational action of teleparallel gravity can be written as follows
\begin{equation}
I=\frac{1}{16\pi G}\int d^{4}x\,|e|\, T \,,
\end{equation}
where $|e|\equiv \det(
e^{\alpha}_{\mu})=\sqrt{-\det(g_{\mu\nu}})$ is the determinant of the vierbein $e^{a}_{\mu}$ which
is equal to $\sqrt{-g}$. Variation of this action with respect
to the vierbeins gives the teleparallel field equations as follows
\begin{equation}
e^{-1}\partial_{\mu}(ee_{i}^{\rho}S^{\,\,\,\,\mu\nu}_{\rho})
-e_{i}^{\lambda}T_{\,\,\,\,\mu\lambda}^{\rho}S^{\,\,\,\,\nu\mu}_{\rho}
+\frac{1}{4}e_{i}^{\nu}T=4\pi G
e_{i}^{\rho}\Theta^{\,\,\nu}_{\rho}\,.
\end{equation}
Now similar to modifying the action of general relativity which $R$
is replaced by a general function $f(R)$, one can replace the
teleparallel action $T$ by a function $f(T)$. Doing this, the
resulting modified field equations are
$$e^{-1}\partial_{\mu}(eS^{\,\,\,\,\mu\nu}_{i})f_{T}(T)
-e^{\lambda}_{i}T^{\rho}_{\,\,\,\,\mu\lambda}S_{\rho}^{\,\,\,\,\nu\mu}f_{T}(T)
$$
\begin{equation}
+S_{i}^{\,\,\,\,\mu\nu}\partial_{\mu}(T)f_{TT}(T)
+\frac{1}{4}e^{\nu}_{i}f(T)=4\pi
G\,e^{\rho}_{i}\,\Theta^{\,\,\nu}_{\rho}\,,
\end{equation}
where $\Theta^{\,\,\,\nu}_{\rho}$ is the energy momentum tensor of
matter, $f_{T}\equiv\frac{df}{dT}$ and
$f_{TT}\equiv\frac{d^{2}f}{dT^{2}}$. We note that we have set the spin connection to zero (as usually it is done in the formulation of $f(T)$ gravity).
 In this case the theory does not have local Lorentz
invariance. In what follows we set $4\pi G=1$.\\

\subsection{Cosmological Considerations}

Now for cosmological considerations we assume a spatially flat FRW
metric as[34,55]
\begin{equation}
e^{\,\,\,\alpha}_{\mu}=diag\,\Big(1,-a(t),-a(t),-a(t)\Big)\,,
\end{equation}
where $a(t)$ is the scale factor of the universe. for Lagrangian
density, we obtain
\begin{equation}
T=-6(\frac{\dot{a}}{a})^{2}=-6H^{2}\,,
\end{equation}
where $H=\frac{\dot{a}}{a}$ is  the Hubble parameter. Therefore, the cosmological
 field equations become
\begin{equation}
12H^{2}f_{T}(T)+f(T)=16\pi G \rho,
\end{equation}
and
\begin{equation}
48H^{2}\dot{H}f_{TT}(T)-4(\dot{H}+3 H^{2})f_{T}(T)-f(T)=16\pi G
\rho,
\end{equation}
where $\rho$ and $p$ are the ordinary matter energy
density and pressure respectively. The conservation equation in this framework is
as follows
\begin{equation}
\dot{\rho}+3H(\rho+p)=0.
\end{equation}
Now, equations (12) and (13) can be rewritten as follows
\begin{equation}
H^{2}=\frac{16\pi G}{3}(\rho+\rho^{(T)})\,,
\end{equation}
and
\begin{equation}
2\dot{H}+3H^{2}=-8\pi G(P+P^{(T)}),
\end{equation}
where $P^{(T)}$ and $ \rho^{(T)}$ are the energy density and pressure of the \emph{torsion fluid} that are defined as
follows, respectively
\begin{equation}
\rho^{(T)}=\frac{2Tf_{T}(T)-f(T)-T}{16\pi G}\,,
\end{equation}
\begin{equation}
P^{(T)}=\frac{4\dot{H}\Big(2Tf_{TT}(T)+f_{T}(T)-1\Big)-\rho^{(T)}}{16\pi
G}.
\end{equation}
Finally, the equation of state parameter of the torsion fluid,
$\omega^{(T)}$, is defined as
\begin{equation}
\omega^{(T)}\equiv\frac{P^{(T)}}{\rho^{(T)}}=
-1+\frac{4\dot{H}\Big(2Tf_{TT}(T)+f_{T}(T)-1\Big)}{2Tf_{T}(T)-f(T)-T}.
\end{equation}
Note that if we consider $f(T)=T+F(T)$, then equations (17) and (19)
can be rewritten as follows
\begin{equation}
\rho^{(T)}=\frac{2TF_{T}-F}{16\pi G}\,,
\end{equation}
\begin{equation}
\omega^{(T)}=-1+\frac{(F-2TF_{T}-T)(F_{T}+2TF_{TT})}{(1+F_{T}+2TF_{TT})(F-2TF_{T})}\,,
\end{equation}
respectively. With these preliminaries, in the next section we study possible
realization of the phantom-like behavior in some observationally viable $f(T)$ models.

\section{Phantom-like Behavior in $f(T)$ Gravity}

To have an effective phantom-like behavior, the effective energy
density must be positive and growing positively with time. Also the
equation of state parameter should be less than $-1$. It has been
shown that modifying the geometric part of the Einstein field
equations alone leads to an effective phantom like behavior while
there is no phantom field in the problem ( see Ref. [44-47]) . Now
we show that modified teleparallel gravity ($f(T)$ gravity) can
realize an effective phantom-like behavior while the null energy
condition for the effective cosmic (torsion) fluid, that is,
$(\rho^{(T)}+P^{(T)}\geq0)$ is respected. We note that as has been
shown in Ref. [53], matter is always stable in these theories. To
see the phantom-like behavior of $f(T)$ gravity, we adopt some
models of $f(T)$ gravity that could be viable through, for instance,
cosmological and solar system tests. In this respect we consider a
scale factor of the type
\begin{equation}
a(t)=\Big(t^{2}+\frac{t_{0}}{1-\nu}\Big)^{\frac{1}{1-\nu}}\,.
\end{equation}
This scale factor has two adjustable parameters, $t_{0}$ and $\nu$
and by assumption $\nu\neq1$. This choice is coming from bouncing
cosmological solutions in the context of string theory with quintom
matter [56]. Note that this choice belongs to the class of scale
factors $a(t)=a_{0}t^{n}$ that are usually adopted in literature.
Based on the definition $1+z=\frac{1}{a(t)}$, with this choice of
the scale factor and equation(11), we find
\begin{equation}
T=\frac{-24\Big(\frac{1}{(1+z)^{1-\nu}}-\frac{t_{0}}{1-\nu}\Big)(1+z)^{2(1-\nu)}}{(1-\nu)^{2}}\,.
\end{equation}
We note that it is easy to show that the scale factor (22) with two adjustable parameters,
$t_{0}$ and $\nu$ is essentially a solution of the field equations. So, in what follows we explore possible realization of
the effective phantom-like behavior with adopted $f(T)$ functions.

\subsection{Phantom-like Effect with $f(T)=T+f_{0}(-T)^{p}$}

In this form of $f(T)$, $f_{0}$ and $p$ are two model's
dimensionless parameters, which only $p$ is an independent parameter
. As we can see, for $p=0$ this type of $f(T)$ gravity can reduce to
$\Lambda CDM$ cosmology, while for $p=\frac{1}{2}$ it reduces to the
Dvali, Gabadadze and Porrati (DGP) braneworld model. Inserting
$f(T)=T+f_{0}(-T)^{p}$ into the first Friedmann equation (12), and
calculating the result for the present time, we obtain $f_{0}$ as
follows [34]

\begin{equation}
f_{0}=\frac{1-\Omega_{m0}}{(6H_{0}^{2})^{p-1}(2p-1)}\,,
\end{equation}
where $\Omega_{m0}$ and $H_{0}$ are the dimensionless density
parameter of the dust matter and the present Hubble parameter
respectively. We note that in order to have a positively
accelerating expansion of the universe, we need those values of $p$
where $p<1$ [57]. Now for this type of $f(T)$ gravity, we find that
(remember that $f(T)\equiv T+F(T)$)
\begin{equation}
F(T)=f_{0}(-T)^{p}\,,
\end{equation}
\begin{equation}
F_{T}=-f_{0}p(-T)^{p-1}\,,
\end{equation}
and
\begin{equation}
F_{TT}=f_{0}p(p-1)(-T)^{p-2}\,.
\end{equation}
By applying equation (20) and (21), we obtain the effective density
and equation of state parameter for this model as
\begin{equation}
\rho^{(T)}=\frac{f_{0}(-T)^{p}(2p-1)}{16\pi G}
\end{equation}

\begin{equation}
\omega^{(T)}=-1+\frac{\Big(pf_{0}(-T)^{p}(1-2p)-pT\Big)}{\Big(pf_{0}(-T)^{p}(1-2p)-T\Big)}.
\end{equation}

Now to see whether this model has an effective phantom-like behavior
or not, we plot the evolution of the effective energy density and
the equation of state parameter versus the redshift, $z$. In this
manner we obtain a suitable amount for the independent parameter $p$
to have an effective phantom-like behavior that this amount is about
$p=0.028$ . In this respect, by using the equations (28) and (29),
figures 1 and 2 are plotted. Also we have used $\Omega_{m0}=0.315$ ,
$H_{0}=67.4km s^{-1}Mpc^{-1}$ and $\nu=1.95$ in all numerical
calculations in the paper. This value $p$ is in agreement with the
results reported in literature. In figures 2, as we can see the
effective equation of state parameter for the present time ($z=0$)
is $\omega_{0}=-1.03$ that this  amount is the favor of the recent
data from Planck2018. However, we should pay attention that in this
form of $f(T)$, the effective equation of state parameter crosses
the cosmological constant (phantom divide) line, $\omega=-1$, from
phantom-like phase to the quintessence-like phase in future. This
behavior is not in the favor of the observational data that show
transition from the quintessence phase to the phantom phase.
Nevertheless, having a positively growing energy density is a
favorable capability of this model.

\begin{figure}[htp]
\begin{center}\includegraphics{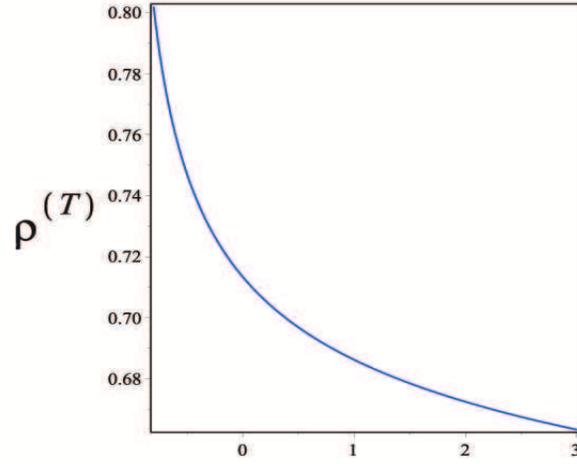} \vspace{7cm}
\end{center}
\caption{\small {Variation of the effective energy density versus
the redshift for $f(T)=T+f_{0}(-T)^{p}$.}}
\end{figure}

\begin{figure}[htp]
\begin{center}\includegraphics{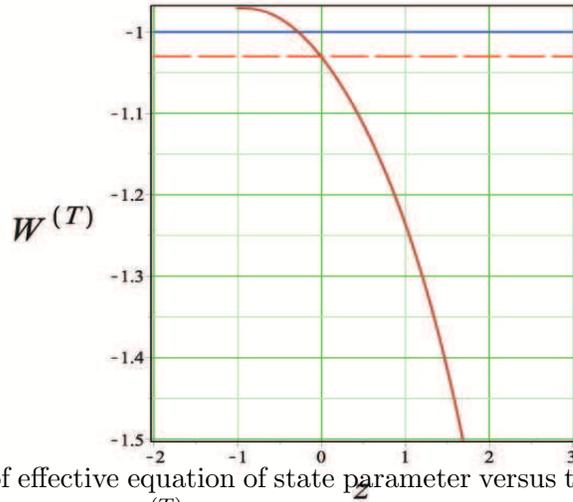} \vspace{7cm}
\end{center}
\caption{\small {Variation of effective equation of state parameter
versus the redshift for $f(T)=T+f_{0}(-T)^{p}$ ,where the dash lines
represent $\omega^{(T)}_{0}=-1.03$ }}
\end{figure}

\subsection{Phantom-like Effect with $f(T)=T+ \xi T(1-e^{\frac{\alpha T_{0}}{T}})$}

In this model $\alpha$ is an independent parameter and
$T_{0}=T(z=0)$ is the current value of the  torsion scaler. In the
limit $\alpha\rightarrow 0$, this model recovers the $\Lambda CDM$
model[34]. Also, by using the modified Friedmann equation, we  find
\begin{equation}
\xi=-\frac{1-\Omega_{m0}}{1-(1-2\alpha)e^{\alpha}}\,.
\end{equation}
In this model, if $\alpha>0$ the universe has a phantom phase (with
$\omega<-1$) and if $\alpha<0$ it has a quintessence phase (with
$\omega>-1$) [48]. For this form of $f(T)$ function we obtain
\begin{equation}
F(T)=\xi T(1-e^{\frac{\alpha T_{0}}{T}})\,,
\end{equation}
\begin{equation}
F_{T}=\frac{\xi\alpha T_{0}}{T}e^{\frac{\alpha
T_{0}}{T}}+\xi(1-e^{\frac{\alpha T_{0}}{T}})\,,
\end{equation}
and
\begin{equation}
F_{TT}=-\frac{\alpha^{2}T_{0}^{2}\xi}{T^{3}}e^{\frac{\alpha
T_{0}}{T}}\,.
\end{equation}
once again, using equations (20) and (21) we acquire $\rho^{(T)} $ and $\omega^{(T)}$ in the present model as
follows
\begin{equation}
\rho^{(T)}=\frac{2\alpha \xi T_{0}e^{\frac{\alpha
T_{0}}{T}}+T\xi(1-e^{\frac{\alpha T_{0}}{T}})}{16\pi G}\,,
\end{equation}
and
\begin{equation}
\omega^{(T)}=-1+\frac{\Big(T\xi(e^{\frac{\alpha
T_{0}}{T}}-1)-2\xi\alpha T_{0}e^{\frac{\alpha
T_{0}}{T}}-T\Big)\Big(\xi-\xi e^{\frac{\alpha
T_{0}}{T}}(1-\frac{\alpha
T_{0}}{T}+\frac{2\alpha^{2}T_{0}^{2}}{T^{2}})\Big)}{\Big(T\xi(e^{\frac{\alpha
T_{0}}{T}}-1)-2\xi\alpha T_{0}e^{\frac{\alpha
T_{0}}{T}}\Big)\Big(\xi-\xi e^{\frac{\alpha
T_{0}}{T}}(1-\frac{\alpha
T_{0}}{T}+\frac{2\alpha^{2}T_{0}^{2}}{T^{2}})+1\Big)}\,,
\end{equation}
respectively. We try to see possible realization of the effective
phantom-like behavior in this model by numerical treatment of these
quantities. We investigated the the behavior of $\rho^{(T)}$  and
$\omega^{(T)}$  versus the redshift $z$ for different choices of the
model parameter $\alpha$. Our analysis shows that the phase of
expansion of the universe (according to the values of $\omega$)
depends on the sign of the parameter $\alpha$. We find that
$\alpha=0.07$  is the acceptable value for realization of an
effective phantom-like behavior in this model. By using the
equations (34) and (35), the evolution of $\rho^{(T)}$ and
$\omega^{(T)}$ versus the redshift are shown respectively in figures
3 and 4. Figure 3 shows an increasing (positively growing) energy
density with the cosmic time (inverse of the redshift). In figure 4,
as we can see with the proper selection of parameters, this model
can display $\omega^{(T)}=-1.03$ for present time. Also figure 4
shows that the effective equation of state stays always less than
$-1$ in this model. Therefore, we note that in this model the
crossing of the phantom divide could not occur and the universe
evermore is in the phantom phase. Similar to the previous case, this
behavior is not in the favor of the observational data that shows
transition from the quintessence phase to the phantom phase.
Therefore, we can conclude that the present model is not
observationally viable.

\begin{figure}[htp]
\begin{center}\includegraphics{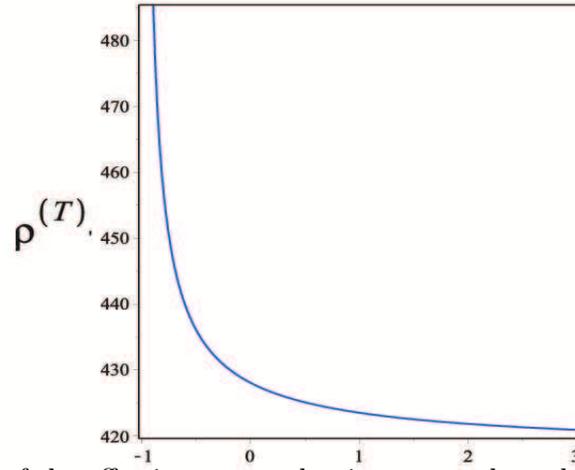} \vspace{7cm}
\end{center}
\caption{\small {Variation of the effective energy density
 versus the redshift for $f(T)=T+ \xi
T(1-e^{\frac{\alpha T_{0}}{T}})$.}}
\end{figure}

\begin{figure}[htp]
\begin{center}\includegraphics{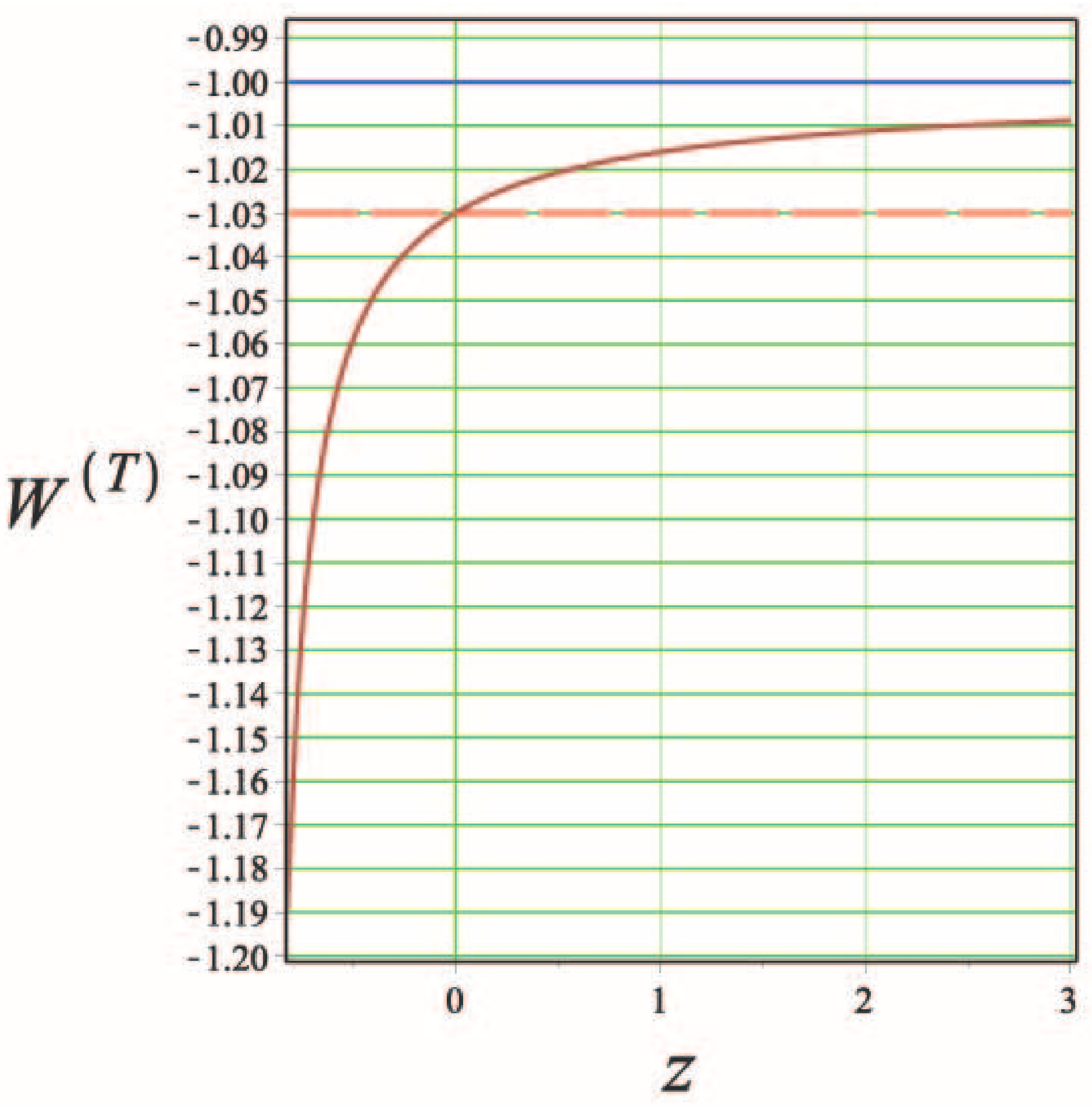} \vspace{7cm}
\end{center}
\caption{\small {Variation of effective equation of state parameter
 versus the redshift for $f(T)=T+ \xi
T(1-e^{\frac{\alpha T_{0}}{T}})$, where the dash lines represent
$\omega^{(T)}_{0}=-1.03$.}}
\end{figure}

\subsection{Phantom-like Effect with $f(T)=T+n\Big(T-Te^{(\frac{\beta T_{0}}{T})}+(-T)^{\beta}\Big) $}

The above two models have been used in the literatures for a variety
of purposes. Possible transition to the phantom phase of the cosmic
expansion has been treated with the mentioned $f(T)$ models in a
context a little different from our approach. Here, and for the sake
of more novelty of the treatment, we consider a combined model that
contains a relatively more general $f(T)$ function. This approach is
introduced in Ref. [48]. The number of free parameters here are more
than the previous cases. Usually with a wider parameter space one
expects to encounter more fine-tuning of the parameter to have
cosmologically viable solutions. But, in the same time it is more
easier, in principle, to find some subspaces of the model parameter
space to fulfill the required expectation. For the sake of
simplicity,we have taken $\xi=f_{0}\equiv n$ and $p=\alpha\equiv
\beta$ .Here $\beta$ and $n$ are two parameters of model that only
$\beta$ is free parameter because $n$ can be obtained as
\begin{equation}
n=-\frac{\Omega_{m0}-1}{((1-2\beta)6H_{0})^{2\beta-2}+\Big(1-(1-2\beta)e^{\beta}\Big)}\,.
\end{equation}
Also for this combined model, we can acquire
\begin{equation}
F(T)=n\Big(T-Te^{(\frac{\beta T_{0}}{T})}+(-T)^{\beta}\Big)\,,
\end{equation}
\begin{equation}
F_{T}=\frac{n\beta T_{0}}{T}e^{\frac{\beta
T_{0}}{T}}+n(1-e^{\frac{\beta T_{0}}{T}})-n\beta(-T)^{\beta-1}\,,
\end{equation}
and
\begin{equation}
F_{TT}=-\frac{n\beta^{2}T_{0}^{2}}{T^{3}}e^{\frac{\beta
T_{0}}{T}}+n\beta(\beta-1)(-T)^{p-2}\,.
\end{equation}
Similar to the former two models, by using equations (20) and (21) we
obtain the effective density and equation of state parameter for
this model respectively as follows
\begin{equation}
\rho^{(T)}=\frac{n\Big((-T)^{\beta}(2\beta-1)+T+e^{\frac{\beta
T_{0}}{T}}(2\beta T_{0}-T)\Big)}{16\pi G}\,,
\end{equation}
and
\begin{equation}
\omega^{(T)}=-1+\Theta \Upsilon\,,
\end{equation}
where we have defined
\begin{equation}
\Theta\equiv\frac{\Big(n(-T)^{\beta}(1-2\beta)+e^{\frac{\beta
T_{0}}{T}}(Tn-2n\beta
T_{0})-T(1+n)\Big)}{\Big((-T)^{n}(1-2\beta)+e^{\frac{\beta
T_{0}}{T}}(T-2\beta T_{0})-T \Big)}
\end{equation}
and
\begin{equation}
\Upsilon\equiv\frac{\Big(1+(\frac{\beta
T_{0}}{T}-\frac{2\beta^{2}T_{0}^{2}}{T^{2}}-1)e^{\frac{\beta
T_{0}}{T}} +(1-2\beta) \beta (-T)^{\beta-1}\Big)
}{\Big(1+n(\frac{\beta
T_{0}}{T}-\frac{2\beta^{2}T_{0}^{2}}{T^{2}}-1)e^{\frac{\beta
T_{0}}{T}}+n+n\beta(1-2\beta)(-T)^{\beta-1} \Big)}\,.
\end{equation}

Finally, to investigate the possible realization of the phantom-like
behavior in this combined model,by using equations (40) and (41), we
plot $\rho^{(T)}$ and $\omega^{(T)}$ versus the redshift for
different values of $\beta$. We observe that the two conditions for
effective phantom mimicry \textbf{a})the effective energy density
must be positive and growing with time \textbf{b}) the equation of
state parameter should be less than $-1$, are satisfied if
$\beta=0.2$. This typical behavior of $\rho_{T}$ and $\omega_{T}$
versus the redshift  are shown in figures 5 and 6 respectively. We
note that unlike the previous two cases, this model is strictly in
the favor of the observational data. As is seen from figure 6, it is
obvious that the effective equation of state parameter can cross the
$\omega=-1$ line at $0.5<z_{cross}<1$ and amount $\omega_{T}$ for
Current time $z=0$  is consistent with the resent observational
data. It means that universe evolves from quintessence-like phase
towards the phantom-like phase. We can conclude that this model is
capable of realizing a suitable dynamical mechanism for getting the
present time cosmic acceleration and transition to a phantom-like
stage in a fascinating manner. Consequently, this model is more
cosmologically viable than the previous two models.

\begin{figure}[htp]
\begin{center}\includegraphics{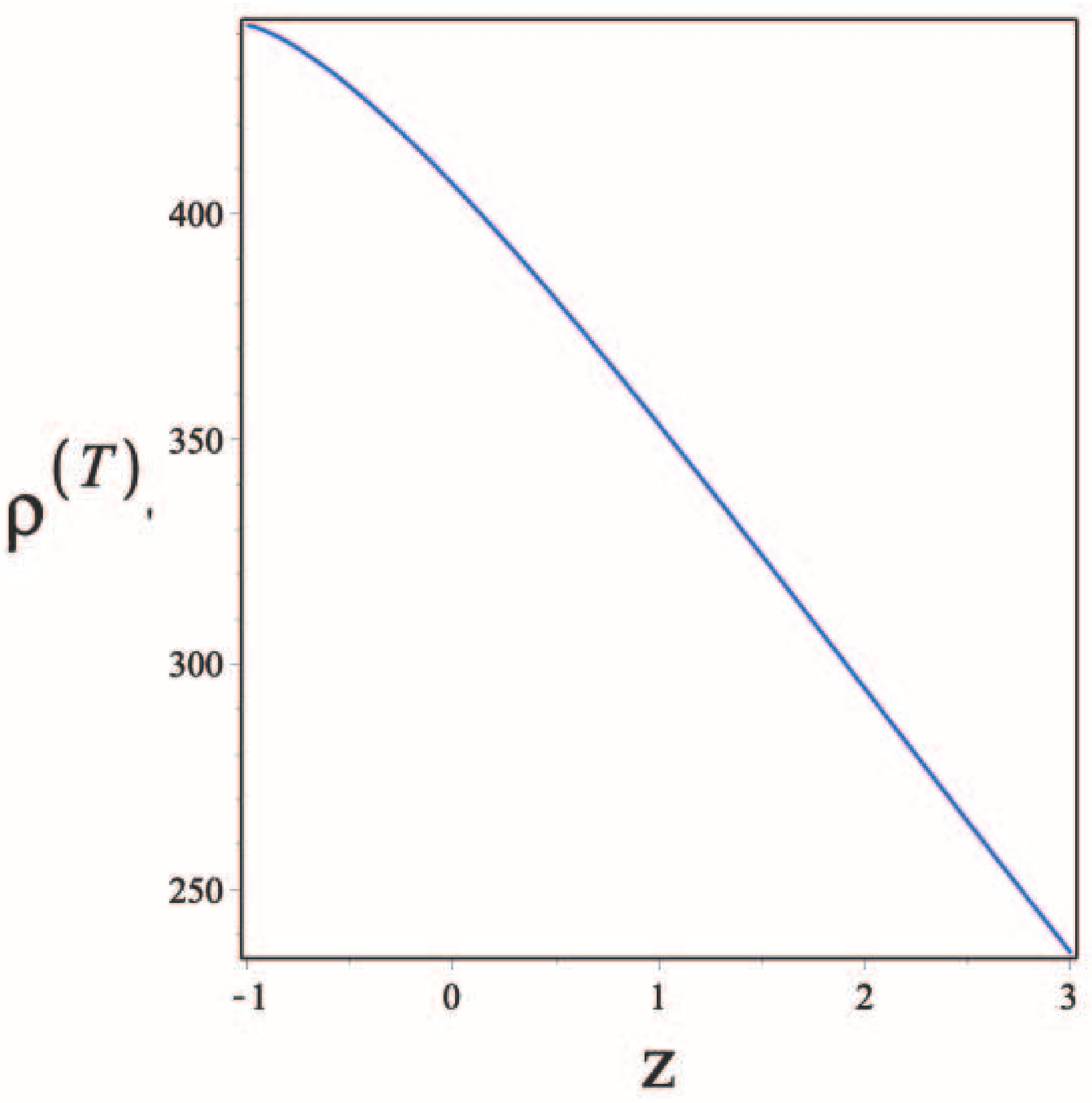} \vspace{7cm}
\end{center}
\caption{\small {Variation of the effective energy density
 versus the redshift for $f(T)=T+n\Big(T-Te^{(\frac{\beta T_{0}}{T})}+(-T)^{\beta}\Big)$.}}
\end{figure}

\begin{figure}[htp]
\begin{center}\includegraphics{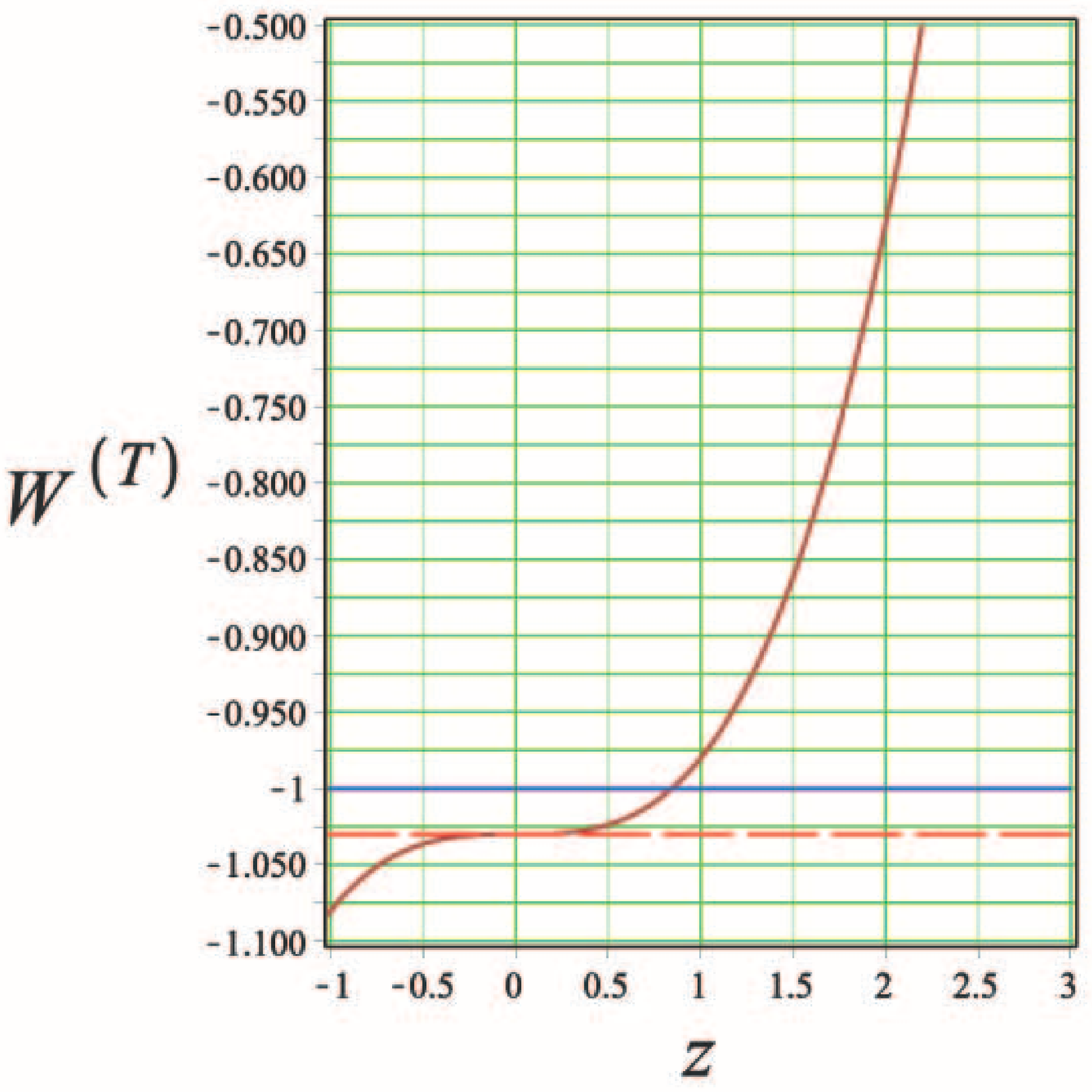} \vspace{7cm}
\end{center}
\caption{\small {Variation of effective equation of state parameter
 versus the redshift for $f(T)=T+n\Big(T-Te^{(\frac{\beta T_{0}}{T})}+(-T)^{\beta}\Big)$, where the dash lines represent
$\omega^{(T)}_{0}=-1.03$.}}
\end{figure}

\section{conclusion}
In this paper we have investigated possible realization of an
effective phantom behavior in some viable $f(T)$ gravity models.
With effective phantom-like behavior we mean an effective energy
density that should be positive and growing positively with time and
also, the equation of state parameter should be less than $-1$. We
have chosen three different models of $ f(T)$ gravity that two of
them are reliable based on previous studies and Our third model is a
combined model. We have shown that model with $f(T)=T+f_{0}(-T)^{p}$
realizes an effective phantom-like behavior if $p=0.028$.As is shown
in figure 2, we obtained  the effective equation of state parameter
for present time about $\omega^{(T)}_{0}=-1.03$ that this amount is
acceptable according to the recent data from Planck2018. But, this
model evolves from effective phantom-like phase towards the
effective quintessence-like phase in future. Thus, this model is not
in the complete favor of observation, since observation shows
transition from quintessence-like to phantom-like phase. We have
presented another $ f(T)$  model, which is $f(T)=T+ \xi
T(1-e^{\frac{\alpha T_{0}}{T}})$. This $ f(T)$  model realizes an
effective phantom like behavior if $\alpha=0.07$. Although,
$\omega^{(T)}_{0}$ is obtained a acceptable value that is shown in
figure 4 ,but this $f(T)$ function is not also in the favor of
observation. Since in this model the crossing of the phantom divide
could not occur and the universe evermore is in the phantom phase .
Finally, we have constructed a $ f(T)$  theory by combining the two
previous models that it only contains  one model parameter $\beta$.
This model realizes an effective phantom-like behavior if
$\beta=0.2$. Unlike the previous two models, this model is
cosmologically more viable, since the equation of state parameter in
this model evolves from effective quintessence-like phase towards an
effective phantom-like phase.As is shown in figure 6, it is obvious
that $\omega^{(T)}$ can cross the $\omega=-1$ line at
$0.5<z_{cross}<1$ and amount $\omega^{(T)}_{0}$ is consistent with
the resent observational data. Therefore, this combined model is in
agreement with the observational data.

\vspace{1cm}

\textbf{Acknowledgment}\\

It is a pleasure to thank Prof. Kourosh Nozari for helpful
discussions and valuable comments.

\vspace{1cm}

\textbf{Data Availability}\\

The data used to support the findings of this study are included
within the article.

\end{document}